\documentclass[twocolumn,showpacs,amsmath,amssymb,bibnotes]{revtex4}
\usepackage{graphicx,epsfig,subfigure,afterpage,bm}

\usepackage{xcolor}
\newcommand{\mybox}[2]{{\color{#1}\fbox{\normalcolor#2}}}

\newcommand{\be}{\begin{equation}}
\newcommand{\ee}{\end{equation}}
\newcommand{\bea}{\begin{eqnarray}}
\newcommand{\eea}{\end{eqnarray}}
\newcommand{\phieff}{\phi}

\newcommand{\etal}{{\it et al.\/}}
\newcommand{\bw}{\begin{widetext}}
\newcommand{\ew}{\end{widetext}}

\newcommand{\Pe}{\rm{Pe}}

\newcommand{\tauo}{\tau_{\rm o}}
\newcommand{\tauc}{\tau_{\rm c}}

\newcommand{\ehat}{\vecv{\hat{e}}}

\newcommand{\vecv}[1]{\bm{{#1}}}
\newcommand{\tens}[1]{\bm{{#1}}}

\begin{document}

\title{Hydrodynamic suppression of phase separation in active suspensions}
\author{Ricard Matas-Navarro$^1$, Ramin Golestanian$^2$, Tanniemola Liverpool$^3$, Suzanne M. Fielding$^1$}
\email{suzanne.fielding@durham.ac.uk}
\affiliation{$^1$Department of Physics, University of Durham, Science Laboratories, South Road, Durham. DH1 3LE}
\affiliation{$^2$Rudolf Peierls Centre for Theoretical Physics, University of Oxford, Oxford, OX1 3NP}
\affiliation{$^3$School of Mathematics, University of Bristol, Clifton, Bristol, BS8 1TW}
\date{\today}
\begin{abstract}
  We simulate with hydrodynamics a suspension of active disks
  squirming through a Newtonian fluid. We explore numerically the full
  range of squirmer area fractions from dilute to close packed and
  show that ``motility induced phase separation'' (MIPS), which was
  recently proposed to arise generically in active matter, and which
  has been seen in simulations of active Brownian disks, is strongly
  suppressed by hydrodynamic interactions.  We give an argument for
  why this should be the case and support it with counterpart
  simulations of active Brownian disks in a parameter regime that
  provides a closer counterpart to hydrodynamic suspensions than in
  previous studies.
\end{abstract}
\maketitle

\section{Introduction}
\label{sec:intro}

``Active matter''~\cite{ActiveSoftReview} comprises internal subunits
that collectively drive the system far from Boltzmann equilibrium by
each individually consuming energy. Biological examples include
actively crosslinked polymeric filaments in the cell
cytoskeleton~\cite{juelicher2007}; cells grouped in living
tissues~\cite{poujade2007}; suspensions of motile
microorganisms~\cite{BergBook,rafai2010}; shoals of fish and flocks
of birds~\cite{ParrishHamnerBook}.  Non-biological examples include
vibrated granular
monolayers~\cite{narayan2007,Tsimring2008,deseigne2010}, and
self-propelled synthetic colloidal
particles~\cite{synthetic,synthetic1}.

Distinct from the more familiar scenario of a passive complex fluid
driven by (say) a global shear flow imposed at the system's
boundaries, in active matter the driving out of equilibrium arises
intrinsically in the active subunits throughout the system's own bulk.
Consequently active materials can spontaneously develop mesoscopic or
macroscopic mechanical stresses and deformations even without driving
or loading from outside. Other exotic and generically emergent
phenomena include swarming, pattern formation, giant number
fluctuations, non-equilibrium ordering and phase separation.
(For a recent review see Ref.~\cite{ActiveSoftReview}.)  These offer
fascinating challenges to fluid dynamicist, rheologist and statistical
physicist alike.

Many active particles are elongated and so have an intrinsic (steric)
tendency to align with each other~\cite{toner1998,toner1995,
  ramaswamy2003,toner2005, baskaran}.  Activity mediated coupling
between these orientational modes and fluctuations in the local number
density then provides a generic mechanism for giant number
fluctuations and phase
separation~\cite{chate,narayan2007,deseigne2010}. The standard
deviation $\Delta N$ in a subregion of material of mean number of
particles $N$ then scales as $N^a$ with $a>1/2$, whereas for passive
systems away from any transition $a=1/2$.

Besides any such tendency for alignment, another generic mechanism for
phase separation was recently put forward in the context of ``run and
tumble" particles, such as some species of motile
bacteria~\cite{tailleur2008a,cates2012a}. These swim in near
straight-line runs at almost constant speed, between intermittent
tumbles in which they suddenly randomize swim direction.  The basic
idea is that particles (a) accumulate where they move more slowly and
(b) move more slowly where crowded. Positive feedback between (a) and
(b) then gives rise to ``motility-induced phase separation''
(MIPS). This idea was recently extended analytically to active
Brownian particles~\cite{TailleurCatesBrownian}, consistent with
simulations showing phase separation in active Brownian
disks~\cite{fily2012a} and spheres~\cite{Redner2012} that indeed
lack any tendency for steric alignment, or for phase separation in the
passive equilibrium case.

To date these simulations lack hydrodynamic interactions, in which
moving particles set up flow fields that influence their neighbours.
But such interactions arise widely in active
matter~\cite{ActiveSoftReview}, and this is fundamentally important
because steady state properties in non-equilibrium systems depend
strongly on dynamics, in contrast to equilibrium states, which depend
only on the underlying free energy. This includes the existence (or
otherwise) of activity-induced phase separation.

The contribution of this work is to show that hydrodynamic
interactions in fact strongly suppress MIPS. Accordingly, MIPS might
not arise as generically in active matter as hitherto suggested. We
show this by simulating with hydrodynamics a suspension of active
disks~\cite{Lighthill1952,Blake1971,Blake2D} that squirm through a
Newtonian fluid.  A closely related model was studied previously by
Pedley and co-workers~\cite{ishikawa2008a,2D3D,ISI:000261251800018}.
We explore the full range of squirmer area fractions from dilute to
close packed, and demonstrate that hydrodynamics causes a key
assumption of the (a) - (b) feedback mechanism outlined above to
fail. It does so by effectively rendering a crucial parameter (defined
below) $\zeta\approx 1$ rather than $\zeta\gg 1$.  To support this, we
further demonstrate suppression of MIPS in active Brownian disks
in the regime $\zeta\approx 1$ more closely analogous to the
squirmers, with MIPS recovered for $\zeta\gg 1$, as explored
previously for active Brownian particles~\cite{fily2012a}.

Being disks, the active particles studied here lack any steric
tendency to align with each other. This choice was made deliberately
in order to exclude a priori the first, non-MIPS mechanism for active
phase separation discussed above. For simulations of active rods with
hydrodynamics, see Ref.~\cite{shelley}.

The paper is structured as follows. In Sec.~\ref{sec:models} we
outline the models to be studied in the rest of paper. In
Sec.~\ref{sec:simulation} we detail our simulation methods. Units and
parameter values are discussed in Sec.~\ref{sec:parameters}, then the
statistical quantities that we measure from the simulations are defined
in Sec.~\ref{sec:measured}. In Sec.~\ref{sec:results} we present our
results, and provide a discussion of them in
Sec.~\ref{sec:discussion}. Finally in Sec.~\ref{sec:conclusions} we
give conclusions and perspectives for future study.

\section{Models}
\label{sec:models}

In this section we outline the models to be studied throughout the
manuscript. We start in Secs.~\ref{sec:introsquirmers}
and~\ref{sec:squirmers} by discussing hydrodynamic squirmers, before
summarising in Sec.~\ref{sec:Brownian} the active Brownian particles
that we shall simulate for comparison with the hydrodynamic case.

\subsection{Squirmers}
\label{sec:introsquirmers}

The model of hydrodynamic swimmers that we shall adopt is based on a
minimal description of microbial propulsion originally put forward by
Lighthill~\cite{Lighthill1952} then further by
Blake~\cite{Blake1971,Blake2D}, and studied extensively by Pedley et
al.~\cite{ishikawa2008a,2D3D,ISI:000261251800018}.  The swimming
particles are assumed to be neutrally buoyant and so force free. Their
size and swimming speed are assumed sufficiently small that the
Reynolds number is negligible and the flow associated with their
swimming is Stokesian. Their size is however assumed large enough that
Brownian motion is negligible (infinite Peclet number).

Swimming through the suspending fluid is achieved by means of an
imposed tangential squirming velocity round the particle surface, but
without changing the particle shape. This is intended to mimic, for
example, locomotion by the beating of many cilia on the surface of a
ciliated microbe. For simplicity the tangential velocity is assumed
time-independent, representing an average over many beating cycles.
Spherical (3D)~\cite{Blake1971} and cylindrical (2D)~\cite{Blake2D}
incarnations of this model have been considered, with axisymmetric
tangential velocity assumed for the spherical case.

The model adopted here is based on the 2D model as put forward by
Blake ({\it i.e.}, infinitely long cylinders). We shall then further
adapt it to the case of a film of disks ({\i.e.}, highly flattened
cylinders) to give a model that has 3D hydrodynamics.

\subsection{Squirming disks}
\label{sec:squirmers}

\begin{figure}[tp]
  \includegraphics[width=9.0cm]{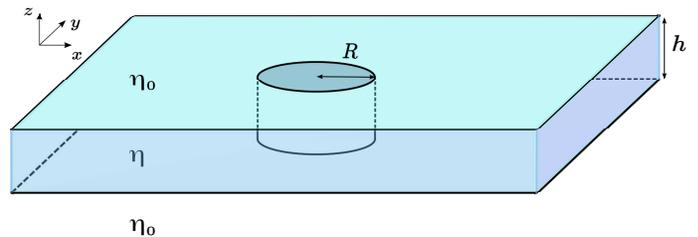}
  \caption{Sketch of the disk geometry.}
\label{fig:drawing}
\end{figure}

\begin{figure}[tp]
  \includegraphics[width=8.5cm]{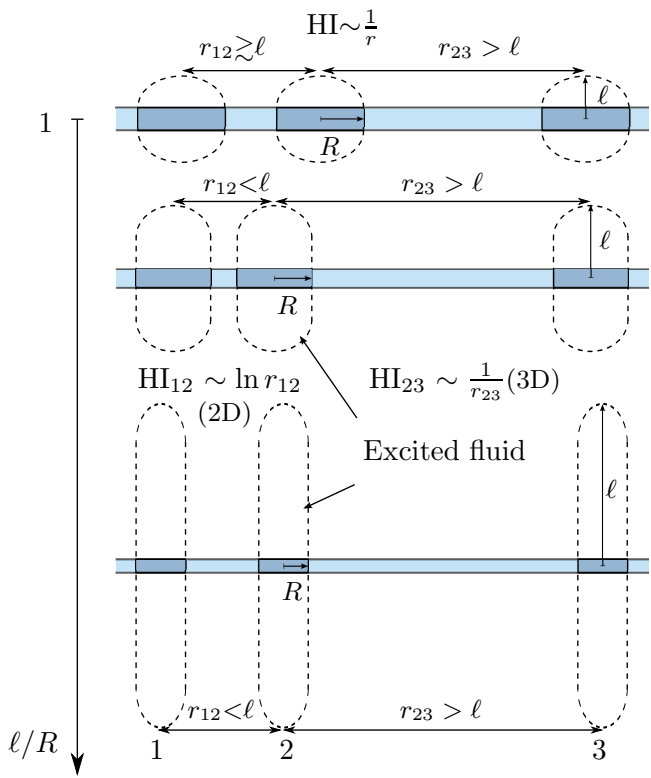}
  \caption{ Schematic of effective description of the hydrodynamic interactions (HI in the figure), i.e. fluid velocities in the film generated by a point force on the disk boundaries.
  The fluid in the bulk is excited by motion in the film and 
  the Saffman length $\ell$ sets the extent of the bulk fluid set in motion. The region of bulk fluid excited by a disk of radius $R$ can be viewed as  a``ghost'' particle and 
  the size of the ghost particles controls the crossover between 2D and 3D nature of the effective interactions~\cite{Saffman,LG1996,LLM2004,LL2010}. 
  When $\ell \leq R$, the effective hydrodynamic interactions (HI) between the disks separated by distance $r$ are 3D-like and always scale to leading order as $1/r$. 
  When $\ell > R$, the effective hydrodynamic interactions between disks depend on the  ratio of the distance between the disks and the Saffman length $\ell$. For a pair of disks with centres at positions $\vecv{r}_i,\vecv{r}_j$, the hydrodynamic interaction (HI) between them is 3D-like if $r_{ij}=| \vecv{r}_i - \vecv{r}_j |  \gg \ell$ and scales like $1/r_{ij}$ while it is 2D-like when $r_{ij}=| \vecv{r}_i - \vecv{r}_j |  \ll \ell$ and scales like $\ln r_{ij}$. }
\label{fig:saff}
\end{figure}

We consider an ensemble of $P$ inertialess disklike particles, each of
radius $R$, actively propelling themselves in the $x-y$ plane of a
horizontal film of an inertialess, incompressible Newtonian fluid of
viscosity $\eta$.  The film has dimensions $L_x,L_y$, with periodic
boundary conditions in the $x$ and $y$ directions.  The disks and the
film each have height $h$, with the top and the bottom of the film at
$z=\pm h/2$.  Semi-infinite volumes of a Newtonian fluid of a lower
viscosity $\eta_0<\eta$ fill the space above and below the film. See
Fig.~\ref{fig:drawing}. This geometry was first considered by Saffman 
and Delbruck in the context of protein diffusion in fluid membranes \cite{Saffman,LG1996,LLM2004,LL2010}.
One can define the Saffman length as $\ell=\eta h/\eta_0$, which quantifies 
the relative significance of the viscous dissipation in the 2D film as compared 
with the 3D dissipation in the bulk. As a simple rule of thumb, one can imagine 
an effective description of the hydrodynamic interactions between the disks, 
by regarding them as objects of typical size set by the Saffman length 
that would interact through 3D bulk viscous hydrodynamics (See Fig.~\ref{fig:drawing} and Fig. ~\ref{fig:saff}).


The disks propel themselves in the plane of the film by means of an
active ``squirming'' motion, achieved by a prescribed tangential
velocity
\be
\label{eqn:slip}
S(\theta-\alpha_p)=B_1\sin(\theta-\alpha_p)+\frac{1}{2}B_2\sin2(\theta-\alpha_p)
\ee
round the disk edges.  We denote by $\beta$ the ratio $B_2/B_1$ of the
first to second modes in this expression. Positive $\beta$ gives
``pullers'', and negative $\beta$ gives ``pushers''.

With these dynamics, a single disk undisturbed by any others has,
in an infinite box, a swim speed $v_0=B_1/2$~\cite{Blake2D}, with an
instantaneous swim direction $\ehat_p=(\cos\alpha_p,\sin\alpha_p)$,
for the $p$th disk.  In a suspension of many disks the actual swim
speeds and directions evolve over time due to hydrodynamic
interactions mediated by the Newtonian fluids surrounding the
particles.

In the fluid above and below the film the velocity and pressure fields
$\vecv{v'}({\bf r},t)$ and $p'({\bf r},t)$ obey the mass balance
condition for incompressible flow
\be
\label{eqn:incomp3D}
0=\nabla \cdot \vecv{v}'({\bf r},t),
\ee
and the Stokes condition of force balance in the limit of creeping flow:
\be
0 = \eta_0 \nabla^2\vecv{v}'({\bf r},t) - \nabla p'({\bf r},t).
\ee
Here ${\bf r}= (x,y,z)$ denotes space and $t$ time.

The fluid in the film has velocity and pressure fields
$\vecv{v}(\vecv{r},t)$ and $p(\vecv{r},t)$ that are taken to obey the
mass balance condition for incompressible flow
\be
\label{eqn:incomp}
0=\nabla_\perp \cdot \vecv{v}(\vecv{r},t),
\ee
and the  force balance condition
\be
\label{eqn:stokes}
0 = \eta \nabla_\perp^2\vecv{v}(\vecv{r},t) - \nabla_\perp p(\vecv{r},t)+\vecv{f}+\vecv{\sigma}^+-\vecv{\sigma}^- \; , 
\ee
where $\vecv{r}=(x,y)$ are position vectors in the plane of the film.
Here we have decomposed the gradient operator as
$\nabla=(\nabla_\perp,\partial_z)$, with $\nabla_\perp$ representing
gradients in the plane of the film.

We recognise Eqn.~\ref{eqn:stokes} as a two dimensional Stokes
equation subject to additional source terms $\vecv{f}$,
$\vecv{\sigma}^\pm$. We choose the term $\vecv{f}$ 
to represent forces round the edge of each disk:
\be
\label{eqn:forces}
\vecv{f}(\vecv{r},t)=\sum_p\vecv{f}_p(\theta_p)\delta(r_p-R).
\ee
In this expression we are summing over the separate polar coordinate
systems $(r_p,\theta_p)$ of the disks, such that for the $p$th term
\be
\vecv{r}=\vecv{R}_p(t)+r_p\cos(\theta_p)\hat{\vecv{x}}+r_p\sin(\theta_p)\hat{\vecv{y}},
\ee
in which $\vecv{R}_p=\vecv{R}_p(t)$ is the position of the centre of
the $p$th disk. These forces are included so as to ensure that the
$p$th disk has at any instant a velocity round its edge
\be
\label{eqn:velocities}
\vecv{v}_p=\vecv{V}_p - R\Omega_p\hat{\vecv{\theta}}_p +
S(\theta-\alpha_p)\hat{\vecv{\theta}}_p,
\ee
comprising solid body translation and rotation, plus the tangential
squirming motion prescribed by the slip velocity function in
Eqn.~\ref{eqn:slip} above. 


The integral properties of the forces are
constrained to ensure zero total force and torque for each disc,
consistent with these particles being swimmers driven by their own
internal dynamics, and not subject to externally imposed force
monopoles.

For mathematical convenience the interior of the discs is also taken to
contain Newtonian fluid obeying Eqns.~\ref{eqn:incomp}
and~\ref{eqn:stokes}. We then simply discard this part of the
solution as we are only interested in the part of the solution outside the discs. 

We note that more general squirmer models could be studied with squirming velocity profiles
prescribed for the interior points at the top and bottom surfaces 
as well as the edges, with the inclusion of a corresponding distribution 
of point forces on the two surfaces to ensure the prescribed 
boundary condition. This would be required for example if one wanted a solid interior of the disk, which our model does not address. 
However we note that even for disks with a solid interior, the model studied here is a good approximation  in the limit of $R \ll \ell$ though it is clearly 
not valid in this case for arbitrary values of $\epsilon=R/\ell$.

Finally, coupling between the film and the surrounding bulk fluid is
achieved by setting
\be
\vecv{\sigma}^\pm=\eta_0\partial_z \vecv{v}'|_{x=\pm h/2},
\ee
at the top and bottom of the disks. In this way, the source terms
$\vecv{\sigma}^\pm$ in Eqn.~\ref{eqn:stokes} represent drag on the
disks by the fluid flow in the bulk just above and below the film.


In the context of active particles, the Peclet number $\Pe$ is defined
as the time taken for a particle thermally to diffuse a distance equal
to its own radius, divided by the time taken for it to swim the same
distance. Here we assume this to be infinite, as in
Refs.~\cite{TailleurCatesBrownian,fily2012a,Redner2012}, suppressing
Brownian motion entirely and considering only the deterministic
hydrodynamics defined above. Physically, this is the relevant limit
for many active suspensions of, {\it e.g.,} swimming bacteria. In
simulations of active Brownian particles the tendency for phase
separation is actually most pronounced in this
limit~\cite{Redner2012}. The fact that we show hydrodynamics to
suppress phase separation in this limit where, without hydrodynamics,
it would be most pronounced, gives strong evidence that hydrodynamics
should further suppress phase separation across the full range of
Peclet number.

As documented in Ref.~\cite{ISI:A1995TH00200013}, the case of strictly
athermal, strictly hard sphere colloids with hydrodynamics is a
singular limit.  To avoid this unphysical pathology we consider the
physically realistic case of slightly soft particles with a pairwise
repulsive force $\vecv{F}_{ij}=-f(a^3-a^2)\vecv{\hat{d}}_{ij}$,
$a=2sR/(d_{ij}-2R)$ for inter-particle separation $d_{ij}<2R(1+s)$
between particles $i$ and $j$.  The effective area fraction is then
$\phi=P\pi R^2(1+s)^2/L_xL_y$.

\subsection{Active Brownian particles}
\label{sec:Brownian}

Following
Refs.~\cite{fily2012a,Bialke2012,McCandlish2012,Redner2012,henkes2011}
we take our active Brownian disks to obey translational dynamics
\be
\vecv{\dot{r}}_i=v_0\ehat_i+\mu \vecv{F}_{ij},
\ee
again with swim speed $v_0$ for a single undisturbed particle.
Their angular dynamics is prescribed by
\be
\label{eqn:angular}
\dot{\theta}_i=n_i(t), \langle n_i(t)n_j(t')\rangle=2\nu_r\delta_{ij}\delta(t-t').
\ee
In contrast to the squirmers, the swim directions of the active
Brownian particles are unaffected by interparticle interactions: they
independently follow Eqn.~\ref{eqn:angular}, regardless of the
frequency or closeness of interparticle encounters.

Use of the word Brownian should be interpreted carefully in this
context. In the dynamics just prescribed, only the angular motion is
stochastic: the translational motion comprises deterministic swimming
with slightly soft repulsive interactions between the particles.  In
any truly thermal system, the angular and translational diffusion
coefficients would be related by a fluctuation dissipation relation.
In contrast, the stochastic angular dynamics used here is not intended
to represent true thermal motion, but a continuous time model of, {\it
  e.g.,} stochastic run-and-tumble events. We use the word Brownian
for consistency with the description of this angularly stochastic
dynamics in the existing literature.

\section{Simulation method}
\label{sec:simulation}

In this section we outline our simulation method for the hydrodynamic
squirmers. (Brownian particles are in comparison far easier to
simulate, using standard methods that we do not discuss here.)  For
simplicity we introduce the method first in the context of the
two-dimensional case $\epsilon=\eta_0 R/\eta h\to 0$ of infinitely
long cylinders propelling themselves in the plane of their cross
section through a Newtonian solvent of viscosity $\eta$ by means of
the slip function $S(\theta)$, which is independent of height $z$
along the cylinder.  (For spherical particles, methods related to ours
can be found in
Refs.~\cite{ISI:000297939200003,ISI:000293478200002,ISI:000261251800018,
  ISI:000253764400067,ISI:000223422300002,ISI:A1995TH00200013,ISI:A1988N090500008}.)
Extension to the three-dimensional case of small but non-zero
$\epsilon$, representing disks in a highly viscous film, can then be
shown to follow by relatively a simple modification.

The basis of the simulation method is to calculate at any timestep,
given the known current positions $\vecv{R}_p$ and prefered swim
directions $\alpha_p$ of all the particles, the forces required in
Eqn.~\ref{eqn:forces} to effect the correct edge slip velocity
functions $S$ for each particle in Eqn.~\ref{eqn:velocities}, subject
to the constraints of zero force and torque on each swimmer.  Emerging
from this calculation at each timestep are then the centre of mass
translational and angular velocities for each particle
$\vecv{V}_p,\Omega_p$ in Eqn.~\ref{eqn:velocities}, which are used to
update the particle positions and swim directions.

To implement this, we define the velocity vector
\be
\label{eqn:velocityVector}
\vecv{U}_{pq}=(v_{prqs},v_{prqc},v_{p\theta qs},v_{p\theta qc})^T
\ee
in which $v_{prqs}$ is the Fourier component in $\sin(q\theta_p)$ of
the radial component of velocity round the edge of the $p$th disk, and
$v_{prqc}$ the corresponding cosine mode. The quantities $v_{p\theta
  qs},v_{p\theta qc}$ are their counterparts for the angular
velocity components. In the same way we define the vector of force
components
\be
\label{eqn:forceVector}
\vecv{F}_{pq}=(f_{prqs},f_{prqc},f_{p\theta qs},f_{p\theta qc})^T.
\ee

To obtain a relation
between (\ref{eqn:velocityVector}) and (\ref{eqn:forceVector}) we start by
taking the curl and plane Fourier transform $(x,y)\to (k_x,k_y)$ of
Eqn.~\ref{eqn:stokes}:
\be
\label{eqn:stokesB}
\eta k^4\psi(\vecv{k})=[\vecv{k}\wedge\vecv{f}(\vecv{k})].\hat{\vecv{z}}.
\ee
(For infinite cylinders, the end-drag terms $\vecv{\sigma}^{\pm}$ are absent.)
Here $\psi$ is the usual stream function, which guarantees that the
incompressibility condition, Eqn.~\ref{eqn:incomp}, is satisfied.

It is then possible exactly to express $\vecv{U}_{pq}$ in terms of the
stream function $\psi(\vecv{k})$, and likewise $\vecv{F}_{pq}$ in
terms of $\vecv{k}\wedge\vecv{f}(\vecv{k})$. Together with
Eqn.~\ref{eqn:stokesB}, this gives
\be
\label{eqn:solve}
\vecv{U}_{pq}=\sum_{p'=1}^{P}\sum_{q'=0}^Q \tens{M}_{qq'}(\vecv{R}_p-\vecv{R}_{p'})\cdot \vecv{F}_{p'q'},
\ee
in which $\tens{M}_{qq'}(\vecv{R}_p-\vecv{R}_{p'})$ is a matrix
propagator that exactly relates the $q'$th Fourier mode of ring forces
for a disk centred at $\vecv{R}_{p'}$ to the $q$th mode of disk edge
velocities of a disk centred at $\vecv{R}_p$. It contains a sum over
$N_k$ plane Fourier modes for each of $k_x$ and $k_y$. For numerical
convenience it is calculated once over a grid of $N_r,N_\theta$
points, then at each timestep of each run looked up by interpolation
as needed.

At each numerical timestep we solve Eqn.~\ref{eqn:solve} given the set
of current disk locations $\vecv{R}_p(t)$ and preferred swimming
angles $\alpha_{p}(t)$, which feature in $\vecv{U}_{pq}$. Prior to
solution, the modes of the imposed squirming function for each disk
are known in $\vecv{U}_{pq}$, while the translational and rotational
velocity of each disk are unknown. Conversely the forces modes are
unknown, apart from the constraints of zero net force and torque for
each disk. After transferring all knowns to the RHS and unknowns to
the LHS, we then invert Eqn.~\ref{eqn:solve} to find the unknowns. The
particle positions and swim angles are then updated as $\vecv{R}_p\to\vecv{R}_p+Dt\vecv{V}_p$, $\alpha_p\to\alpha_p+Dt\Omega_p$, with timestep $Dt$.

In the limit $Q\to\infty$, $N_k\to\infty$, $N_r\to\infty$,
$N_\theta\to\infty$ this gives an {\em exact} solution of the {\em
  full} hydrodynamics of the two-dimensional case of squirming
cylinders in a periodic box of dimensions $L_x, L_y$. Emergent effects
include the net propulsion of each disk, power law far-field
interactions, and lubrication forces in near-field. In numerical
practice, $Q, N_k, N_r, N_\theta$ are of course finite, representing
the maximum number of modes used in the simulation. Repeating the
simulation for progressively larger $Q, N_k, N_r, N_\theta$ ensures
convergence on these parameters to the desired limit $Q\to\infty$,
$N_k\to\infty$, $N_r\to\infty$, $N_\theta\to\infty$.


In the two dimensional limit just discussed, a single such cylinder
subject to a net external force in the plane of its cross section
would suffer Stokes' paradox.  In practice, of course, any such
catastrophe is avoided here by the fact that each squirmer is free of
any net force monopole, experiencing only higher multipoles of force
generated by its own internal squirming dynamics.

Nonetheless, purely 2D hydrodynamics is still to be treated with
caution.  Accordingly, we also consider the three-dimensional case of
small but non-zero $\epsilon=R /\ell$, which corresponds to
disklike particles moving in a highly viscous film, surrounded by a
bulk fluid of much lower viscosity on either side. In this case the
hydrodynamic interactions between the disks can be shown to follow, in
the ``Saffman'' limit~\cite{Saffman} of small $\epsilon$, 
by means of a simple modification to the propagator in Eqn.~\ref{eqn:solve}. 
The main effect of this is to modulate the far-field power law index by
one (which would in fact remove Stokes paradox even if our particles
were subject to external forces).  Accordingly, once the 2D (cylinder)
code has been written, the 3D (disks in a film) case follows by
a straightforward modification of the propagator.


\section{Parameter values and units}
\label{sec:parameters}

The squirmer model has nine parameters: the number of disks $P$; the
disk radius $R$; the box size $L_x=\L_y=L$ in the $x-y$ plane; the
single particle swim speed $v_0$; the amplitude $f$ of the repulsive
potential; the range $s$ of the repulsive potential; the Newtonian
viscosity $\eta$; the ratio $\beta=B_2/B_1$; and the Saffmann
parameter $\epsilon=\eta_0 R/\eta h$.

The active Brownian model has eight parameters, including $P, R, L, v_0,
f$ and $s$, as for the squirmers.  We then further have the drag
coefficient $\mu$ (analogous to the Newtonian viscosity for the
squirmers), and the ratio $\zeta=v_0\sqrt{P}/\nu_rL$ of the prescribed
time of decorrelation of swim direction and the characteristic time
interval between particle collisions.

In each case the number of independent parameters is reduced by three
by choosing units of length in which the box size $L=1$; of time in
which the single particle swim speed $v_0=1$; and mass in which
$\eta=1$ (squirmers) or $\mu=1$ (Brownian).

For both squirming and Brownian dynamics we then take the number of
particles $P=128$ or $P=256$, heavily constrained by computational
cost for the squirmers, with these $P$ values being comparable to that
achieved in other squirmer studies~\cite{ishikawa2008a}. Much larger
values of $P$ are achievable for the Brownian disks, but we present
results only for the same values of $P$ as for the squirmers to ensure
as direct a comparison as possible. We have checked that all the
phenomena reported here are robust to changing between $P=128$ and
$P=256$.

For both squirming and Brownian dynamics the repulsive potential has
$f=1,s=0.1$ to ensure the dimensionless parameter $v_0/\mu f$ (and its
counterpart in the squirming case) prescribing the small extent to
which particles explore the interparticle repulsive potential is
comparable to that in Ref.~\cite{fily2012a}, giving almost hard disks.
We have checked that our results are robust to reasonable variations
in the value of $s$.

We have found the ratio $\beta\equiv B_2/B_1=0$ for the squirmers to
be unimportant for the phase separation phenomenon of interest in this
work. Accordingly, we show results below only for the case $\beta=0$.
However we have verified that our result showing suppression of phase
separation holds across a wide range of values of $\beta=-\infty, -5,
-1, 0, +1, +5, +\infty$ (with $-\infty$ and $+\infty$ actually
equivalent for this model).

This leaves just two dimensionless parameters to be explored
numerically in each case.  For the squirmers we have the
area fraction $\phi=P\pi R^2(1+s)^2/L^2$, and the Saffmann parameter
$\epsilon=\eta_0 R/\eta h$. For the Brownian disks we have the area
fraction $\phi$, as for the squirmers, and the ratio
$\zeta=v_0\sqrt{P}/\nu_rL$ of the prescribed time of decorrelation of
swim direction and the characteristic time between particle
collisions.

\section{Measured quantities}
\label{sec:measured}

Here we define the various statistical quantities that we shall report
in the results section below.
\begin{itemize}
\item

The swim speed
\be
v=\frac{1}{PT}\int_0^T\sum_{p=1}^P|v_p(t)|,
\ee
averaged over the ensemble of particles $p=1\cdots P$ and over a time
interval $t=0\to T$ large enough to get good statistics.  Note that
this measure of speed we choose to adopt differs slightly from the
choice in Ref.~\cite{fily2012a}, extracted from the early-time
ballistic regime of the mean squared particle displacement, which one
can show analytically is equivalent to
$\left[\frac{1}{PT}\int_0^T\sum_i|v_i(t)|^2\right]^{1/2}$.

\item The scaled number fluctuations $\delta_N=\Delta N/N^{1/2}$ in a
  region of the sample with average number of particles $N$. To
  measure this we divide the sample into $b=1\cdots B=P/N$ equally
  sized boxes, define the time-averages $\bar{n_b},\bar{n^2_b}$ of the
  number and squared-number of particles in the $b$th box, then report
  \be
  \delta_N=\frac{1}{B\sqrt{N}}\sum_{b=1}^B\sqrt{\bar{n^2_b}-\bar{n_b}^2}.
  \ee

\item The characteristic time $\tauo$ for the decorrelation of
  particle swim direction, defined as the time interval $\Delta
  t=\tauo$ for the correlation function
\be
P(\Delta  t)=\frac{1}{PT}\int_0^Tdt\sum_p\vecv{e}_p(t+\Delta t).\vecv{e}_p(t)
\ee
  to fall to $1/e$.

\item The time between inter-particle collisions, for which we find a
  reasonable definition to be
\be
\tauc=\pi R/(v_0-\tilde{v}),
\ee
in which
\be
\tilde{v}=\frac{1}{PT}\int_0^T\sum_p \vecv{v}_p(t).\vecv{e}_p(t)
\ee
is a measure of the extent to which a particle manages actually to
attain its full velocity in its attempted swim direction. The degree
to which this differs from the free swim speed is a measure of the
degree to which scattering occurs.  We checked by direct observation
that $\tauc$ provides a reliable measure at low area fraction.  At
high area fraction this comparison is more difficult to carry out,
because particles slither round each other continuously.

\end{itemize}

\section{Results}
\label{sec:results}

\begin{figure}[htp]
  \includegraphics[width=9.0cm]{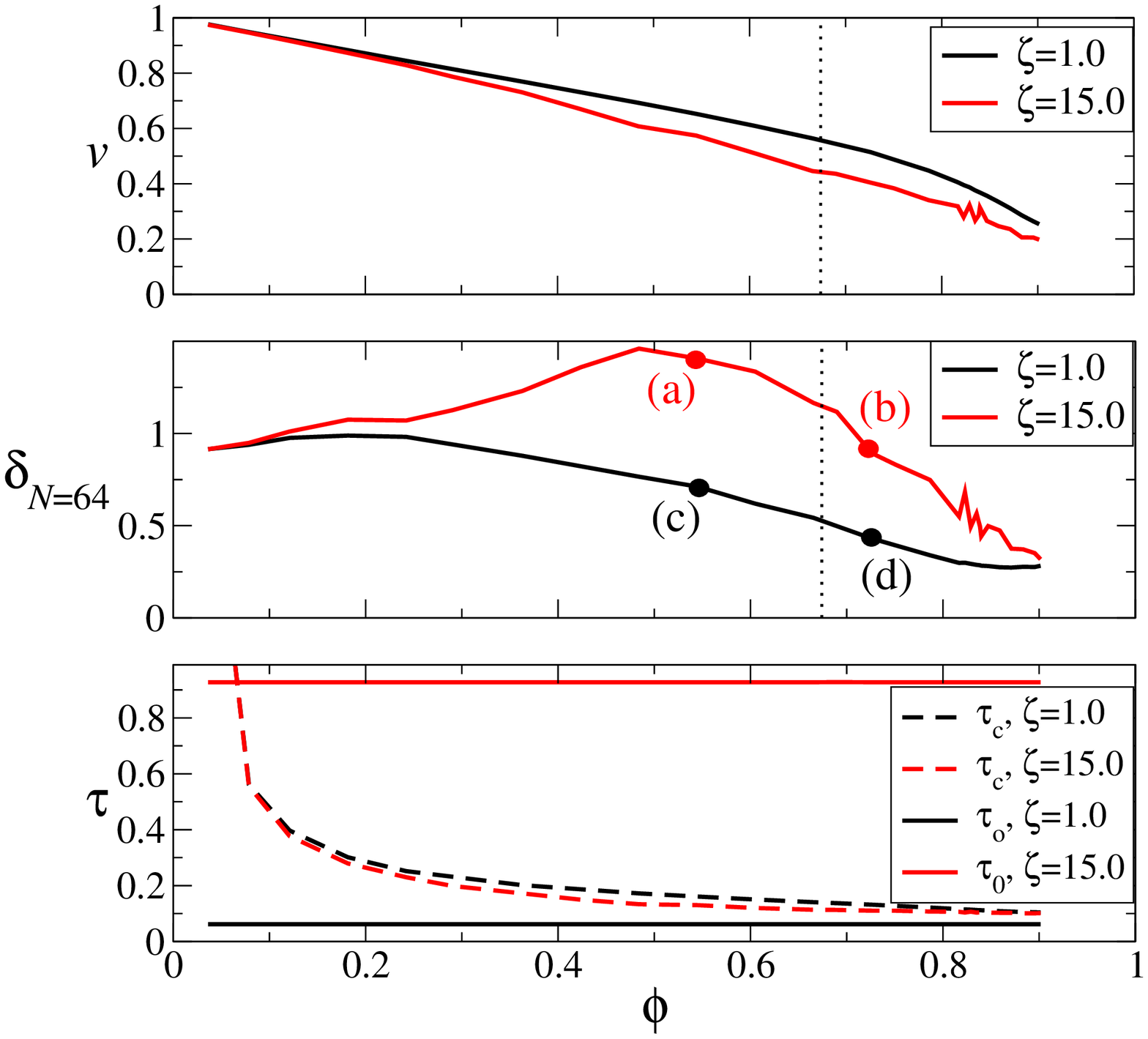}
  \caption{Active Brownian particles. Top: average particle speed as a function of average particle area fraction. Expected spinodal shown by dotted line. Middle: scaled number fluctuations. Bottom: characteristic time $\tauo$ for reorientation of particle swim direction; and inter-particle collision time $\tauc$.}
\label{fig:Brownian}
\end{figure}

\begin{figure}[htp]
\subfigure[ $\zeta=15.0,\phieff=0.5445$]{
\mybox{red}{\fbox{\includegraphics*[width=3.65cm, bb = 62 82 458 450]{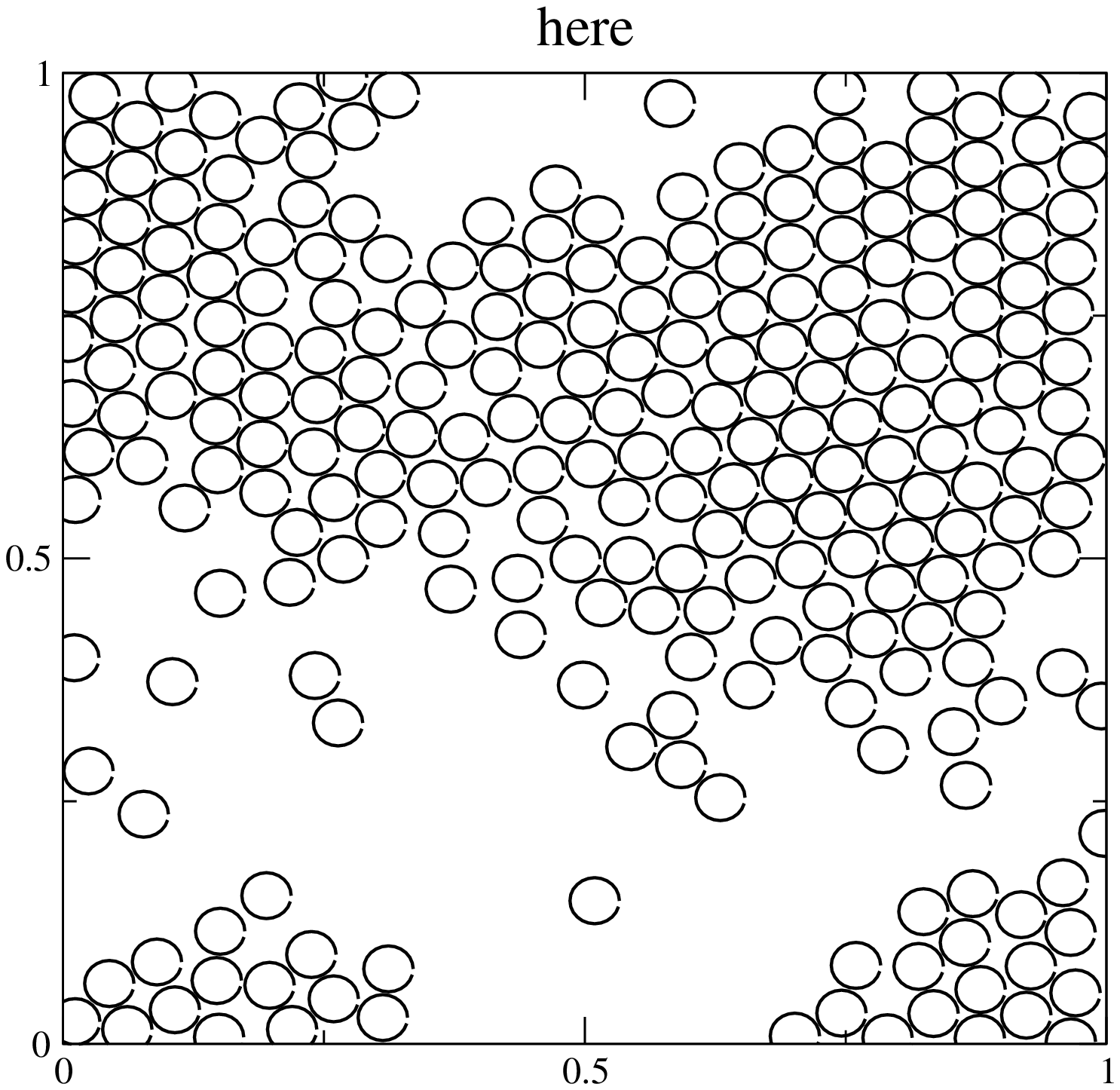}}}
}
\subfigure[ $\zeta=15.0,\phieff=0.726$]{
\mybox{red}{\fbox{\includegraphics*[width=3.65cm, bb = 62 82 458 450]{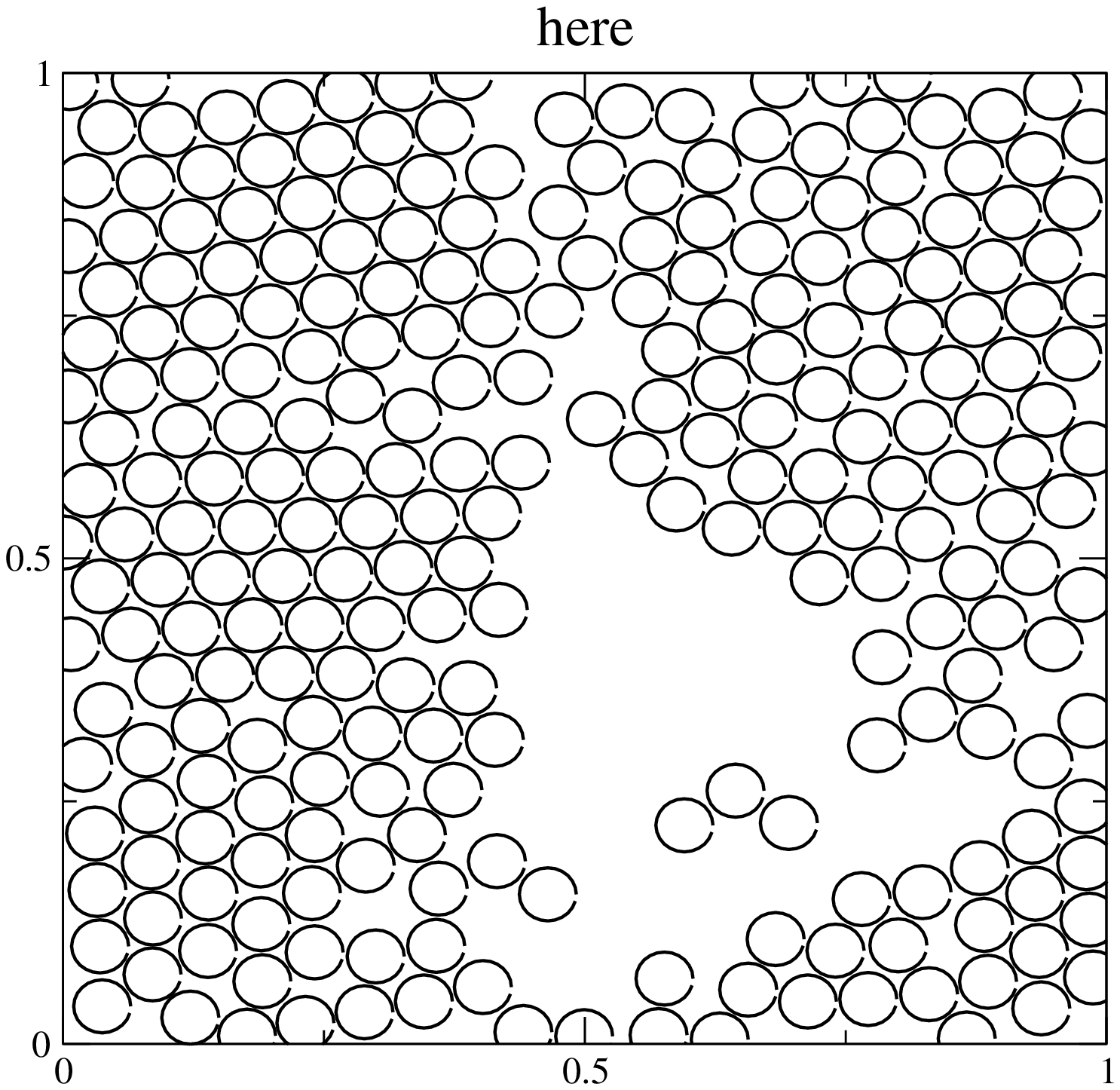}}}
}
\subfigure[ $\zeta=1.0,\phieff=0.5445$]{
  \includegraphics*[width=3.75cm, bb = 57 77 463 455]{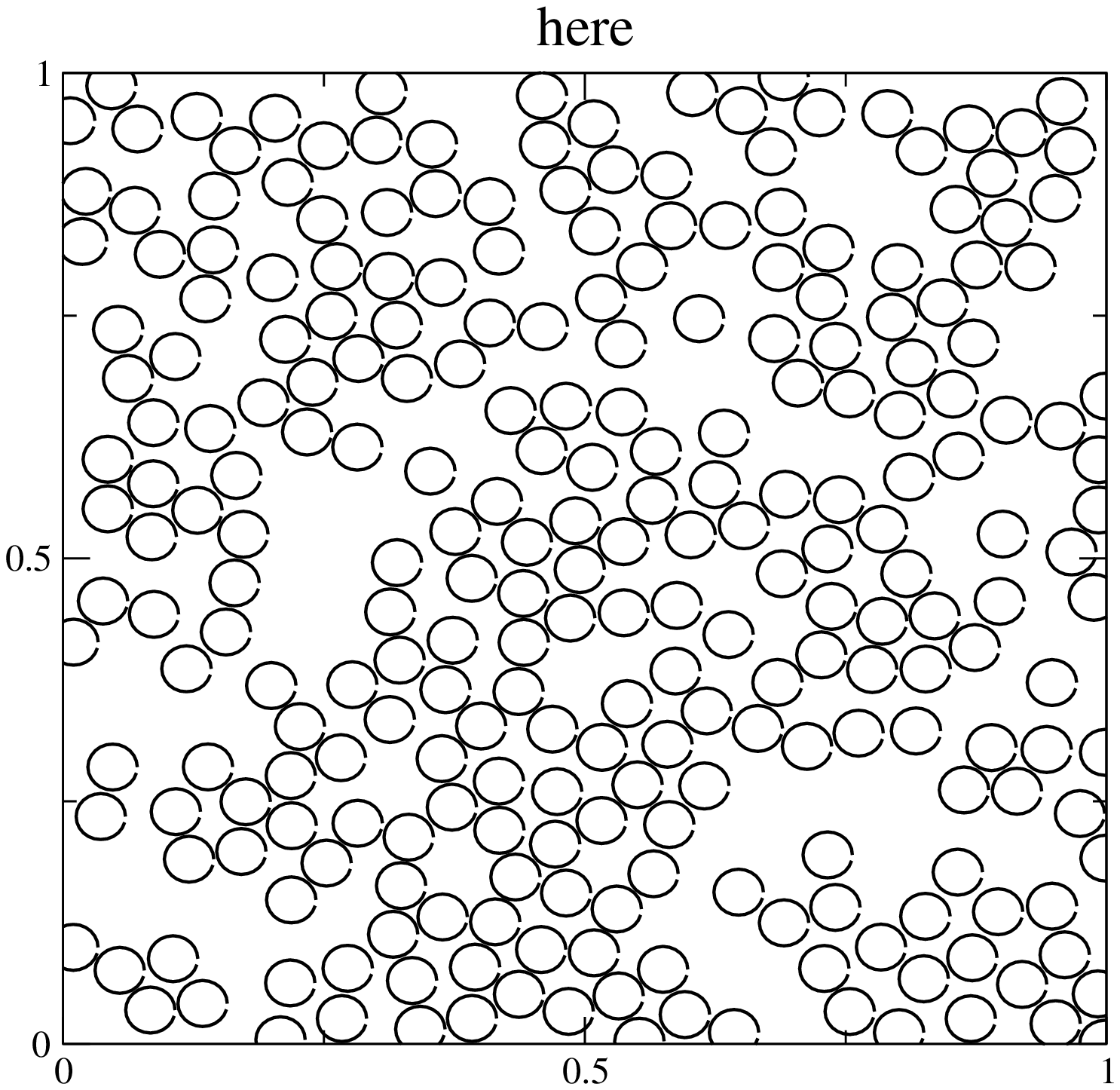}
}
\subfigure[ $\zeta=1.0,\phieff=0.726$]{
  \includegraphics*[width=3.75cm, bb = 57 77 463 455]{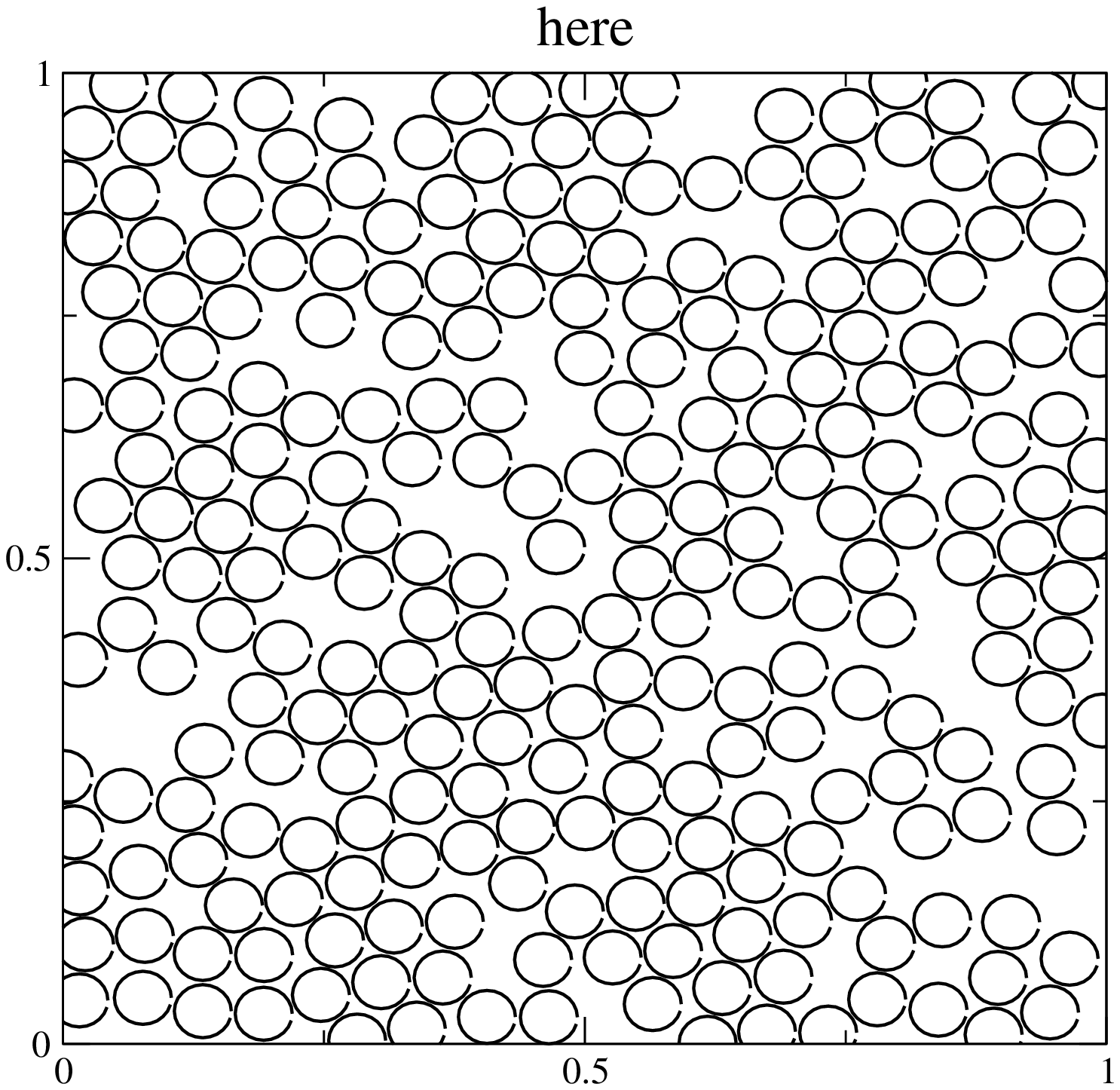}
}
\caption{Snapshots for active Brownian particles, corresponding to circles (a)-(d) in Fig.~\ref{fig:Brownian}.}
\label{fig:BrownianSnap}
\end{figure}

We explore first active Brownian disks in
the regime where the particle swim direction is slow to decorrelate,
$\zeta\gg 1$. We do this to establish a point of contact with earlier
simulation studies~\cite{fily2012a} of active Brownian disks, which
were performed in this regime, to provide a context in which to
discuss our new results below.

As seen in Fig.~\ref{fig:Brownian} (top), and correspondingly in Fig.
3 of Ref.~\cite{fily2012a} (though there with a much larger $P=10^4$),
the particle swim speed $v$ decreases strongly with area fraction
$\phi$.  Accordingly this system is a good candidate for MIPS via the
(a)-(b) feedback mechanism discussed above. The expected spinodal,
according to the instability criterion $d\log v/d\log\phi<-1$ of
Ref.~\cite{tailleur2008a,cates2012a}, is located at the vertical
dotted line in Fig.~\ref{fig:Brownian}.

Phase separation is indeed observed: see the top two snapshots of
Fig.~\ref{fig:BrownianSnap}, and correspondingly Fig.~1 of
Ref.~\cite{fily2012a}. This is reflected also in enhanced number
fluctuations: our plot of $\delta_N(N)$ for each $\phi$ (not shown)
shows all the same features as Fig. 3 of Ref.~\cite{henkes2011},
though obviously cuts off sooner at high $N$ due to our smaller number
of simulated particles $P$.  Reporting in Fig.~\ref{fig:Brownian}
(middle) the single value $\delta_{N=P/9}$, close to the peak of
$\delta_N(N)$ for this value of $P$, clearly indicates separation with binodal
onset around $\phi=0.25$.

\begin{figure}[htp]
  \includegraphics[width=8.5cm]{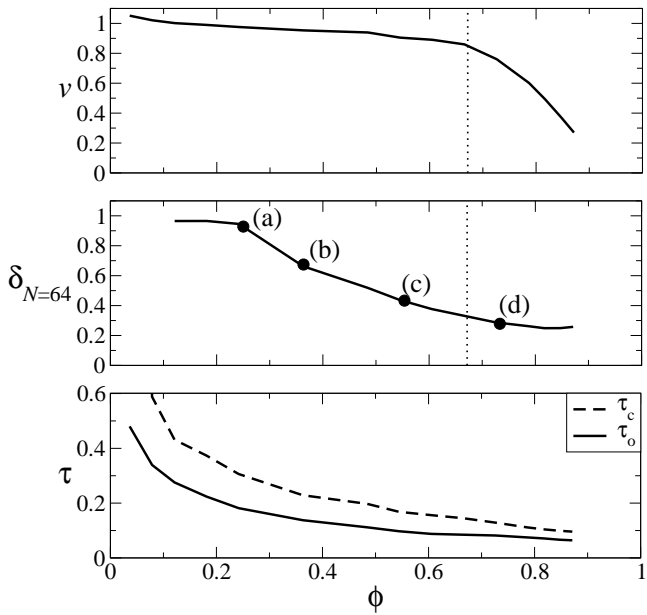}
\caption{Hydrodynamic squirmers (cylinders). Top: average particle speed as a function of average particle area fraction. Expected spinodal shown by dotted line. Middle: scaled number fluctuations. Bottom: characteristic time $\tauo$ for reorientation of particle swim direction; and inter-particle collision time $\tauc$.}
\label{fig:squirmers}
\end{figure}

\begin{figure}[htp]
\subfigure[ $\phieff=0.242$]{
\includegraphics*[width=3.75cm, bb = 57 77 463 455]{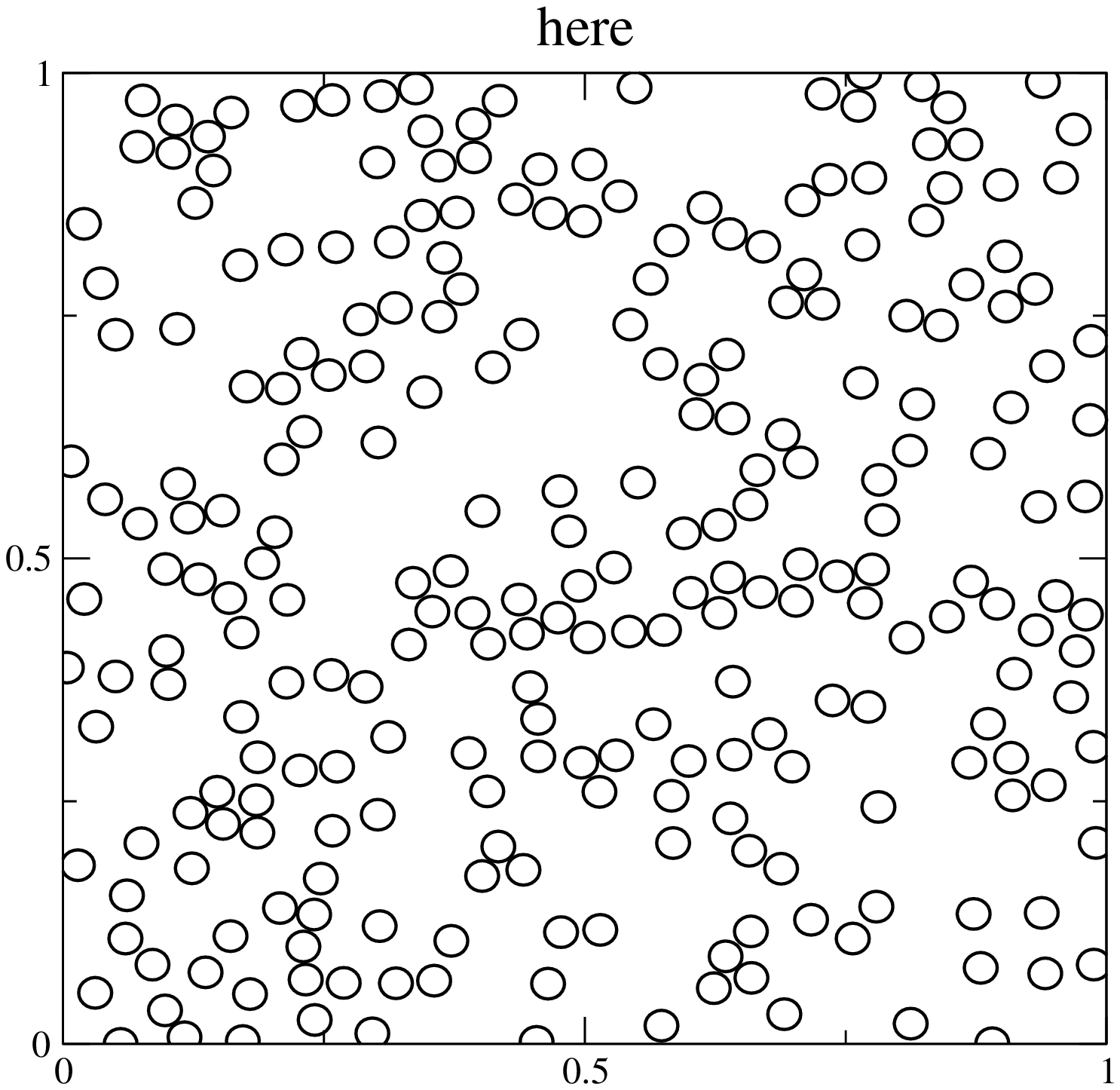}
}
\subfigure[ $\phieff=0.363$]{
\includegraphics*[width=3.75cm, bb = 57 77 463 455]{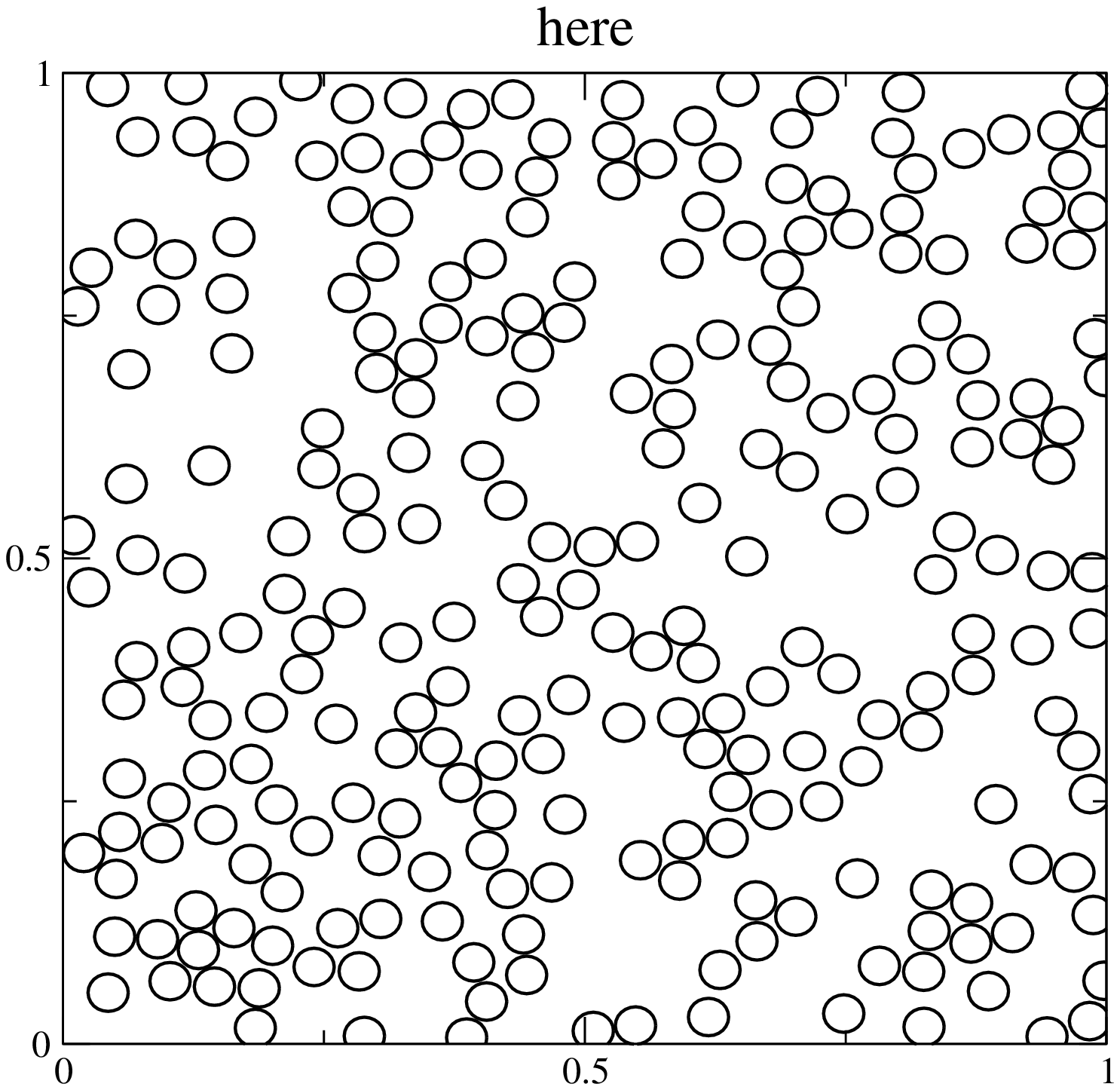}
}
\subfigure[ $\phieff=0.5445$]{
\includegraphics*[width=3.75cm, bb = 57 77 463 455]{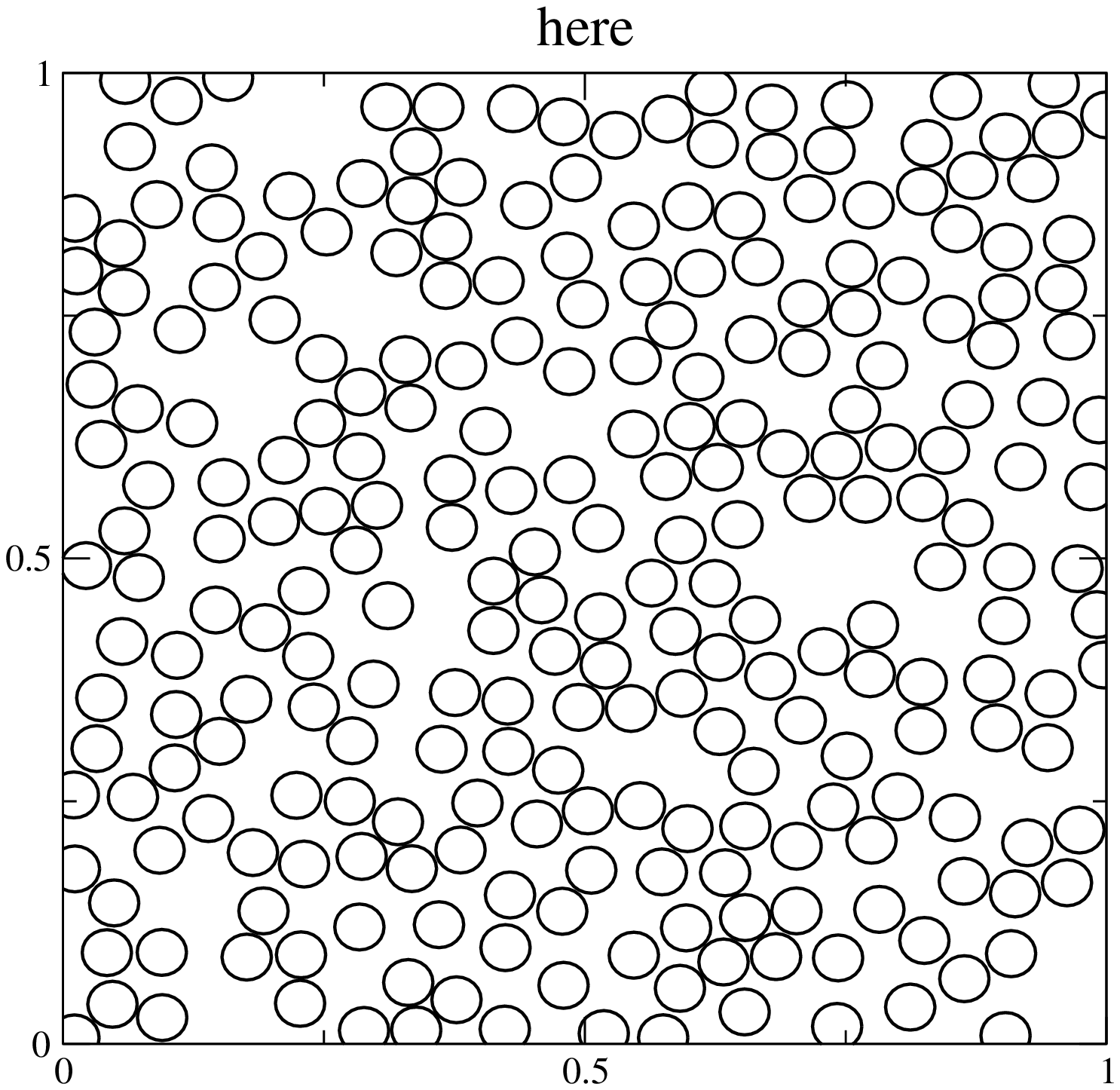}
}
\subfigure[ $\phieff=0.7865$]{
\includegraphics*[width=3.75cm, bb = 57 77 463 455]{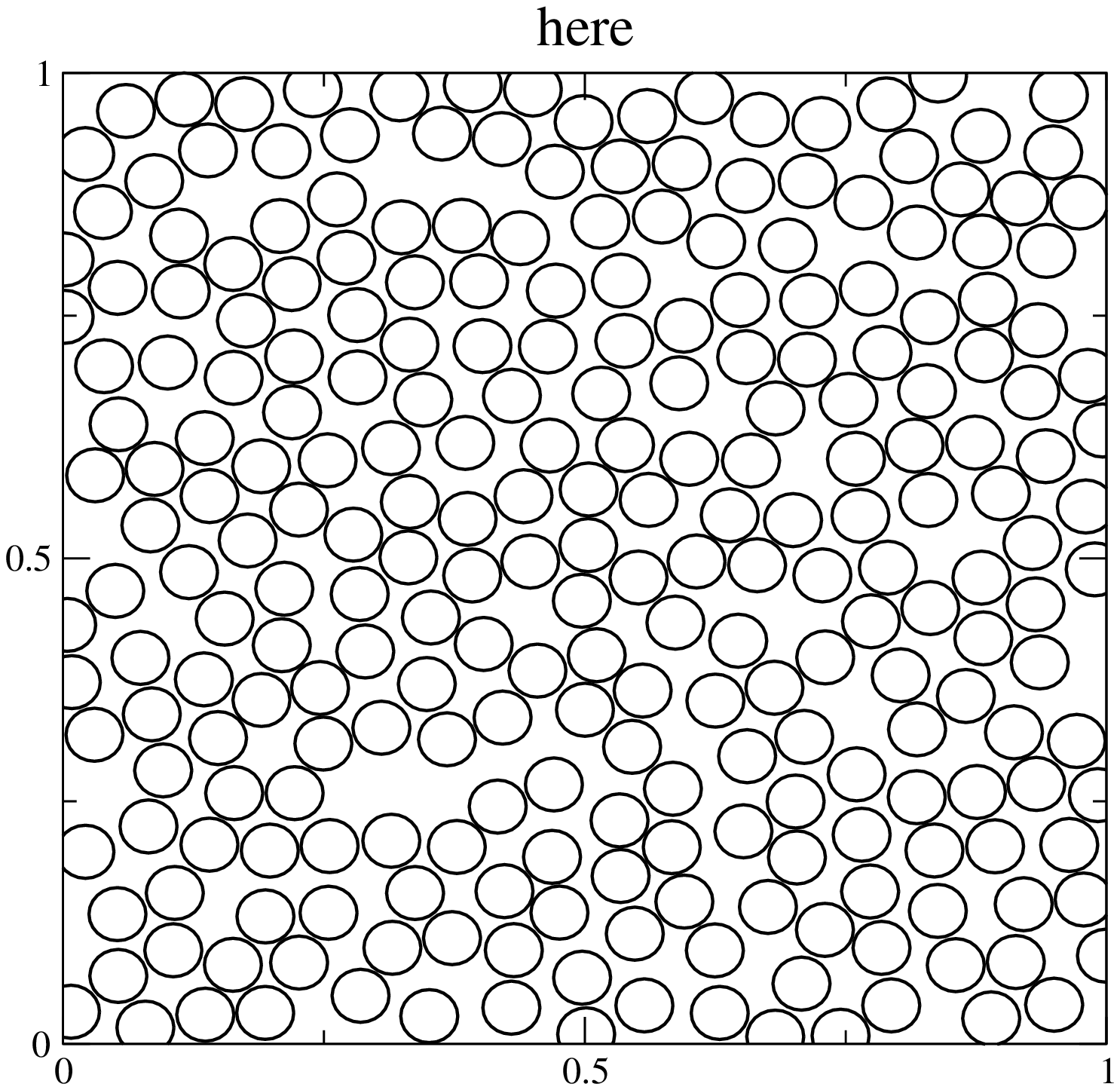}
}
\caption{Snapshots for hydrodynamic squirmers, corresponding to circles (a)-(d) in Fig.~\ref{fig:squirmers}.}
\label{fig:squirmerSnap}
\end{figure}

Consider now the hydrodynamic squirmers. We start for simplicity with
the 2D case $\epsilon=0$ (cylinders), our results for which are shown
in Fig.~\ref{fig:squirmers}. As for the Brownian disks the ensemble
average swim speed declines strongly with particle area fraction,
suggesting that MIPS should again occur via the (a)-(b) feedback
outlined above, with spinodal onset at the vertical dotted line
according to the criterion $d\log v/d\log \phi<-1$ of
Ref.~\cite{tailleur2008a}.  Remarkably, however, we find no evidence
for bulk phase separation in the 2D squirmers: see the snapshots of
Fig.~\ref{fig:squirmerSnap} and the correspondingly suppressed number
fluctuations $\delta_{N=P/9}$ in Fig.~\ref{fig:squirmers} (middle
panel). We do however find a tendency to form small string-like
clusters at low area fractions, as reported previously by other
authors~\cite{ishikawa2008a}.

\begin{figure}[htp]
  \includegraphics[width=8.5cm]{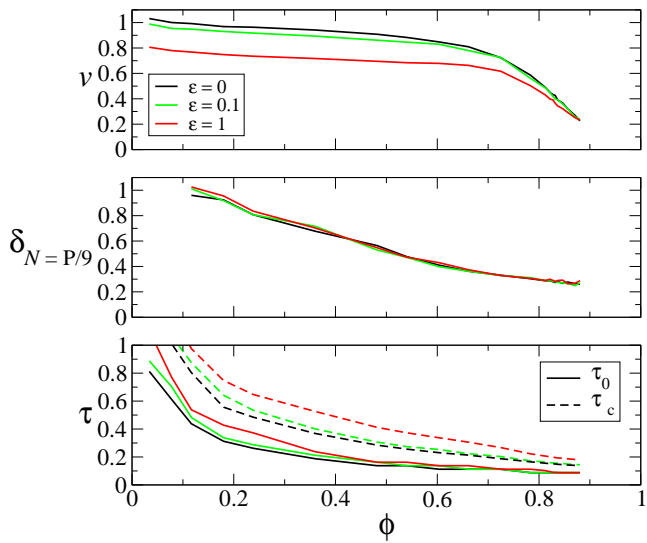}
  \caption{Hydrodynamic squirmers (disks). Top: average particle speed
    as a function of average particle area fraction. Middle: scaled
    number fluctuations. Bottom: characteristic time $\tauo$ for
    reorientation of particle swim direction; and inter-particle
    collision time $\tauc$.}
\label{fig:squirmDiscs}
\end{figure}

To check that this result is not particular to the case of 2D
hydrodynamics discussed so far, $\epsilon=0$, we show counterpart
results for the case of small but non zero $\epsilon$ in
Fig.~\ref{fig:squirmDiscs}. As discussed above, this corresponds to the 3D
hydrodynamics of disks squirming in a film of highly viscous fluid,
surrounded above and below by a bulk fluid of much lower viscosity.
As can be seen, there is no evidence of phase separation in this case
either. Corresponding state snapshots (not shown) closely resemble
those in Fig.~\ref{fig:squirmerSnap} for the cylinders, again showing
no evidence of phase separation.

All these squirmer simulations were run for long times (typically
$t_{\rm max}=20.0$, which in our units is the time taken for any free
squirmer to cross the entire simulation box 20 times). In each case,
we checked the run was long enough to ensure that all the statistical
quantities defined above had convincingly reached a steady state.

\section{Discussion}
\label{sec:discussion}

To understand how hydrodynamics might cause this
suppression of motility-induced phase separation (MIPS) we now
revisit the original argument for MIPS, as first put forward in the
context of run and tumble dynamics and later generalised to other
systems.

Model run and tumble particles move in a series of straight line runs
at a constant speed $v_0$, when dilute, between tumbles in which they
rapidly reassign their swim direction.  These tumbles occur randomly
with a typical intertumble time interval $\alpha^{-1}$.  The argument
for phase separation stemming from this motility is as follows.  (a)
It can be shown that the local particle area fraction in any small
region of fluid scales inversely with the local swim speed $v$.
(Although intuitively reasonable, this is in fact a strongly
non-equilibrium effect stemming from activity.)  (b) It is assumed
that between tumbles particles move more slowly in regions of high
volume fraction, being impeded by crowding, rendering the swim-speed a
decreasing function of volume fraction $v(\phi)<v_0$.  Positive
feedback between (a) and (b) gives phase separation, onset with
spinodal instability from an initially homogeneous state when $d\log
v/d\log \phi < -1$. This argument extends to active Brownian particles
of angular diffusivity $\nu_r$, with an exact mapping $\alpha\to
(d-1)\nu_r$ in $d$ dimensions~\cite{TailleurCatesBrownian}.

Consider carefully part (b) of this argument.
In order meaningfully to define an intertumble swim speed
$v(\phi)<v_0$ that is reduced by crowding, each particle must
encounter many other particles between tumbles: the characteristic
intertumble time $\alpha^{-1}$ must be large compared to the
characteristic timescale $\tau_{\rm c}$ between collisions.
Transcribing this reasoning to Brownian particles, the angular
diffusion time must be small compared to $\tauc$.  Generalizing still
further, we propose that regardless of the underlying microscopic
dynamics -- run and tumble, Brownian or hydrodynamic -- the decay time
$\tauo$ of the swim-direction autocorrelation function must greatly
exceed the inter-particle collision timescale $\tauc$ if part (b) of
the argument is to hold and MIPS is to occur.

With this reasoning in mind consider again Fig.~\ref{fig:Brownian} for
Brownian disks in the regime $\zeta\gg 1$ discussed above.  The bottom
panel of this figure shows the decay time $\tauo$ of the
autocorrelation function of particle swim direction, together with the
typical time $\tauc$ between particle collisions.  Consistent with the
externally imposed angular diffusivity $\nu_{\rm r}$ being small in
these runs (large $\zeta=15$), $\tauo$ is a relatively large constant
across all $\phi$. Except in the very dilute limit each particle
encounters many others during the time it takes to alter its swim
direction: $\tauc\ll\tauo$.  Between angular reorientation events,
then, each particle properly samples the ensemble average reduced swim
speed $v(\phi) <v_0$: part (b) of the feedback loop holds and MIPS can
indeed occur, as observed. Indeed, this mean-field nature of $v(\phi)$
was noted in Ref.~\cite{TailleurCatesBrownian}.

In contrast, for the squirmers we are not at liberty to prescribe from
the outset the timescale $\tauo$ of decorrelation of particle swim
direction: instead this emerges naturally as a result of hydrodynamic
interactions between the particles, which are a priori unknown.  (Put
differently, there is no externally tunable $\zeta$ for these
particles.) Indeed, two-squirmer studies show that with hydrodynamics
each scattering event (``collision'') typically results in an $O(1)$
change in swim-direction for each particle involved.  Accordingly, in
these many-squirmer simulations we expect $\tauo\approx\tauc$. This is
indeed observed: Fig.~\ref{fig:squirmers} (bottom panel).  Squirmers
will therefore be unable properly to sample the reduced $v(\phi)<v_0$
between reorientation events: part (b) of the feedback argument fails,
and MIPS is suppressed.

To support this argument we finally revisit the Brownian disks, but
now imposing a smaller $\zeta=1.0$ so that the particles reorient their
swim directions much more quickly than in the large $\zeta$ case
discussed above, giving dynamics more closely akin to the squirmers
($\tauo\approx\tauc$, Fig.~\ref{fig:Brownian}, bottom).  Phase
separation is then indeed strongly suppressed: see the bottom two
snapshots of Fig.~\ref{fig:BrownianSnap}, and the significantly
reduced number fluctuations of Fig.~\ref{fig:Brownian} (middle panel).
This is despite the ensemble average velocity $v(\phi)$ still
decreasing strongly enough for this $\zeta=1.0$ to satisfy the condition
for spinodal onset. Put simply, particles that rapidly reorient do not
have time to sample this mean field $v(\phi)$ in between reorientation
events.

Our findings are consistent with an earlier comment in Ref.~\cite{HO},
that a tendency to order was suppressed by hydrodynamics in a
suspension of rod-like swimmers; with experiments demonstrating highly
cooperative dynamics and mesoscale turbulence in living fluids,
without associated evidence for phase separation~\cite{wensink}; and
with experiments~\cite{cisneros} demonstrating that concentrated
bacterial suspensions are dominated by a competition between short
range lubrication and steric effects.

Despite the absence of true bulk phase separation for the squirmers,
and for the Brownian disks with $\tauo\approx\tauc$, some particle
clustering is nonetheless still apparent in
Figs.~\ref{fig:BrownianSnap}c,d and Fig.~\ref{fig:squirmerSnap}a-d.
Whether this clustering can be interpreted in terms of a nearby but
suppressed MIPS, which could therefore be a generic feature of active
matter with hydrodynamics, remains an open question.

\section{Conclusions}
\label{sec:conclusions}

We have simulated with hydrodynamics a suspension of active squirming
disks across the full range of area fractions from dilute to close
packed. In doing so we have shown that hydrodynamic interactions
strongly suppress motility induced phase separation.  These findings
should apply generically to active systems in which hydrodynamics are
important, and in which the effective particle velocity depends on
volume fraction via collisions impeding particle motion.  Obvious
exceptions include systems in which $v$ depends on $\phi$ instead by
chemically mediated mechanisms (such as quorum
sensing~\cite{Bassler}), allowing MIPS to arise at relatively low
volume fractions, and/or systems in which the particles reside in (or
on the surface of) a
gel~\cite{bio}.

Open for further study is the degree to which particle elongation
might force a correlation of particle orientations, potentially
restoring phase separation for sufficiently elongated particles, and
so possibly even bringing an understanding of phase separation in
active rodlike suspensions~\cite{chate,narayan2007,deseigne2010} into
the framework proposed here. Also unresolved remains the effect of
dimensionality of particle packing and hydrodynamic propagator: 3D
packings with 3D hydrodynamics do not appear to phase
separate~\cite{ishikawa2008a}, nor do the 2D packings with 2D
hydrodynamics (cylinders) or 3D hydrodynamics (disks) reported
here. However studies of 2D packings of spherical particles with 3D
hydrodynamics~\cite{2D3D} did report separation, though at an area
fraction $\phi=0.1$ that seems very low for MIPS to be implicated.
This issue of dimensionality deserves careful future attention. It
also remains an open challenge to address the clustering observed
experimentally in synthetic colloids~\cite{synthetic1}, noting that
(true) interparticle attractions cannot be entirely eliminated
experimentally. Such questions notwithstanding, the mechanism proposed
here is expected to arise widely in active matter, and particularly in
active colloids, which form the focus of intense current experimental
interest~\cite{synthetic,synthetic1}.

{\it Note added} We note that after the submission of the first
version of our manuscript
~\cite{arxiv}, we learned of a
related work by A. Z\"{o}ttl and H. Stark 
~\cite{stark} studying a squirmer suspension confined between plates
separated by a distance comparable to the swimmer size.  What
differentiates our work from theirs is the fact that hydrodynamic
interactions under confinement are effectively screened with a
screening length that is set by the gap size. In our study of an
extended and {\em unconfined} system, hydrodynamic interactions are
always long-ranged, while the thin film geometry allows us to tune the
effective dimensionality of the long-ranged hydrodynamic interactions
to make a smooth crossover from 2D to 3D behaviour.

{\it Acknowledgements} We thank Mike Cates, Paul Chaikin and Oliver
Harlen for discussions. The research leading to these results has
received funding from the European Research Council under the European
Union's Seventh Framework Programme (FP7/2007-2013) / ERC Grant
agreement no.  279365.  We thank the KITP for hospitality while some
of this work was done: this research was supported in part by the
National Science Foundation under Grant No. NSF PHY11-25915.


\end{document}